\documentclass[12pt]{article}

\usepackage{epsfig}
\usepackage{amsbsy}
\usepackage{amsmath}

\begin{document}

\title{Relations between classical phase-space distributions and Wigner functions for multiparticle production processes.}
\author{K.Zalewski
\\ M.Smoluchowski Institute of Physics
\\ Jagellonian University, Cracow\footnote{Address: Reymonta 4, 30 059 Krakow,
Poland, e-mail: zalewski@th.if.uj.edu.pl This work has been partly supported by
the Polish
 Ministry of Education and Science grant 1P03B 045 29(2005-2008)}
\\ and\\ Institute of Nuclear Physics, Cracow}
\maketitle

\begin{abstract}
The effects of interpreting classical phase space distributions as Wigner
functions, which is common in models of multiparticle production, are
discussed. The temperature for the classical description is always higher than
that for its Wigner function interpretation. A rough estimate shows that the
corresponding correction is proportional to $R^{-2}$, where $R$ is the radius
of the interaction region, and that it is negligible for heavy ion scattering,
but at the few percent level for $e^+e^-$ annihilations.
\end{abstract}
\noindent PACS numbers 25.75.Gz, 13.65.+i \\Bose-Einstein correlations,
interaction region determination.
\section{Introduction}

Much work is being done on the femtoscopy of the interaction regions. One of
the main problems is to find the space-time distribution of the set of the
freeze-out points, i.e. of the points where the hadrons are finally freed. This
is known to depend on the momenta of the particles, which significantly
complicates the problem. For reviews of the work in this field see e.g.
\cite{WIH}, \cite{CSO}, \cite{LIS}.

One can use several functions to describe the geometry of the interaction
region in connection with the corresponding momentum distribution of the final
state particles. The simplest is the classical phase space distribution for the
particles at freeze-out $F(\textbf{p},\textbf{x},t)$. Many models provide just
that. This is immediately seen when classical equations are being used as e.g.
the Euler equations from hydrodynamics or the classical Boltzmann kinetic
equation. For a discussion of a number of cascade models from this point of
view see \cite{AIC}. The classical approach is intuitive and most useful to get
a general picture of the situation. In principle it contradicts quantum
mechanics, because it is not possible to ascribe to a particle simultaneously a
position in space and a momentum. In practice, however, often the quantum
corrections are not very significant.

Another possibility is to use the Wigner function $W(\textbf{p},\textbf{x},t)$.
This is well defined in quantum mechanics. Its relations to the density
matrices in the momentum and coordinate representations are

\begin{eqnarray}
W(\textbf{K},\textbf{X},t) &=& \int\!\!\frac{d^3q}{(2\pi)^3}\;\rho(\textbf{K},\textbf{q},t)e^{i\textbf{qX}}, \\
\label{wigrho} W(\textbf{K},\textbf{X},t) &=&
\int\!\!\frac{d^3y}{(2\pi)^3}\;\tilde{\rho}(\textbf{X},\textbf{y},t)e^{-i\textbf{Ky}
},
\end{eqnarray}
where

\begin{equation}\label{}
  \textbf{K} = \frac{1}{2}(\textbf{p} + \textbf{p}'),\quad \textbf{q} = \textbf{p} - \textbf{p}',
  \quad \textbf{X} = \frac{1}{2}(\textbf{x}+\textbf{x}'),\quad \textbf{y} = \textbf{x} - \textbf{x}'.
\end{equation}
In a rigorously understood sense \cite{TAT},\cite{HIL} the Wigner function is
the best quantum-mechanical replacement for the classical phase-space density.
Heisenberg's uncertainty principle is reflected by the inequality

\begin{equation}\label{wigbou}
  |W(\textbf{p},\textbf{x})| \leq \pi^{-3},
\end{equation}
which follows from the definition of the Wigner function. Wigner functions
integrated over momenta give the correct space distributions and integrated
over the space give the correct momentum distributions. The quantum mechanical
averages of the type $\langle x^mp_x^n\rangle$ cannot, in general, be reliably
calculated using the classical product with the Wigner function as weight,
because they depend on the ordering of the noncommuting operators $x$ and
$p_x$. The averages calculated with the Wigner function always give the quantum
mechanical average for the symmetrized (Weyl's ordering) product. E.g.

\begin{equation}\label{}
  \int\!\!dxdp\;W(p,x)p_x^2x = \frac{1}{4}\langle \hat{p}_x^2\hat{x} + 2\hat{p}_x\hat{x}\hat{p}_x +
  \hat{x}\hat{p}_x^2\rangle,
\end{equation}
Here and in the following the hats are used to distinguish operators from the
corresponding classical quantities.  The most annoying feature of the Wigner
function is that only in very exceptional cases it is nonnegative. In fact, for
pure states the Wigner function is nowhere negative if and only if the
corresponding wave function is a Gaussian \cite{HUD}. Fortunately, for mixed
states this implies that any average over Gaussians satisfying (\ref{wigbou})
can be a Wigner function,which is enough to reproduce almost any shape,
provided there are no peaks violating the bound (\ref{wigbou}). According to
the class of models described in the following section, in order to describe
the multiparticle momentum distributions it is necessary to know the single
particle density matrix $\rho_1(\textbf{p};\textbf{p}')$. As seen from
(\ref{wigrho}), this can be calculated when the Wigner function is known. It
cannot, however, be obtained directly from the classical phase space
distribution $F(\textbf{p},\textbf{x},t)$. Therefore, models which yield the
classical density usually tacitly assume that it is sufficiently similar to the
corresponding Wigner function to replace it in formula (\ref{wigrho}).

The purpose of the present paper is to study the relation between the functions
$F(\textbf{p},\textbf{x},t)$ and $W(\textbf{p},\textbf{x},t)$. Our analysis
suggests that this replacement is legitimate for heavy ion scattering, but
overestimates the temperature of the system by several per cent for $e^+e^-$
annihilations.

Still another possibility is to use the emission function \cite{SHU},
\cite{PRA1}, \cite{PRA2}, related to the density matrix by the relation

\begin{equation}\label{}
  \rho(\textbf{p},\textbf{p}') = N\int\!\!d^4X\;S(K,X)e^{iqX},
\end{equation}
where $q$ and $X$ are four-vectors and N is a constant factor. This formula is
applicable for times after freeze-out has been completed. Then, in the
interaction representation, the density matrix does not depend on time any
more. The emission function is particularly convenient when the time spread of
the freeze-out process is of interest, In the present paper only simultaneous
freeze-out will be considered, so the emission function will not be needed.

\section{Simplifying assumptions}

The multiparticle system just after freeze-out is in some complicated, highly
correlated state. Therefore, in order to deal with it, it is necessary to
introduce approximations. The simplest would be to neglect all the
correlations. Then the diagonal elements of an $n$-particle density matrix in
the momentum representation, which is what one needs to get the $n$-particle
momentum distribution, would be given by the formula

\begin{equation}\label{}
  \rho_{nu}(\textbf{p}_1,\ldots,\textbf{p}_n;\textbf{p}_1,\ldots,\textbf{p}_n) =
  \prod_{j=1}^n \rho_1(\textbf{p}_j;\textbf{p}_j),
\end{equation}
where $u$ in the subscript stands for uncorrelated. In this approximation,
however, for $n$ identical mesons there are no Bose-Einstein correlations.
Since the Bose-Einstein correlations yield important information about the
particle distributions in coordinate space, a better approximation must be
used. The next choice \cite{KAR} (for reviews see e.g. \cite{WIH}, \cite{CSO},
\cite{LIS}) is to introduce proper symmetrization over the momenta of identical
particles. Then for $n$ identical mesons

\begin{equation}\label{karczm}
   \rho_{n}(\textbf{p}_1,\ldots,\textbf{p}_n;\textbf{p}_1,\ldots,\textbf{p}_n) =
  C_n\sum_P\prod_{j=1}^n \rho_1(\textbf{p}_j;\textbf{p}_{Pj}).
\end{equation}
The summation is over all the permutations of the second arguments of $\rho_1$.
Symmetrizing also over the first arguments would just produce a constant factor
$n!$, so there is no point in doing it.  The normalization constant $C_n$ is
now necessary to ensure the proper normalization of $\rho_n$. With this choice,
the single-particle and two-particle momentum distributions are

\begin{eqnarray}\label{}
  P(\textbf{p}) &=& \overline{C}_1\rho_1(\textbf{p};\textbf{p}),\\
  P(\textbf{p}_1,\textbf{p}_2) &=& \overline{C}_2(P(\textbf{p}_1)P(\textbf{p}_2) + |\rho_1(\textbf{p}_1;\textbf{p}_2)|^2),
\end{eqnarray}
where $\overline{C}_n$ are normalizing constants and the hermiticity of the
density matrix $\rho_1$ has been used. Ansatz (\ref{karczm}) leaves out the
final state interactions. For a study of e.g. resonance production this would
be unacceptable, but for analyses of the interaction regions the available
methods of removing the effects of final state interactions from the data are
good enough \cite{LIS} and Ansatz (\ref{karczm}) is widely used.

Let us assume further that, freeze-out for all the particles happens instantly
and simultaneously at some time $t=0$. With this assumption the emission
function reduces to $\delta(t)W(\textbf{p},\textbf{x})$ multiplied by a
normalizing constant. Thus, there is no need to introduce an emission function
besides the Wigner function, which greatly simplifies the discussion. Moreover,
using the interaction representation one has, for $t \geq 0$, time independent
density matrices and consequently time-independent Wigner functions. This
assumption corresponds to a crude approximation. It would be better (cf. e.g.
\cite{LIS}) to assume that for each particle at its freeze-out its proper time
$\tau$ has some fixed value, common for all the particles. Then, however, the
problem of comparison would become much more difficult.

Finally, we assume that the particle density at freeze-out is given by the
canonical distribution corresponding to some non-relativistic problem for
noninteracting particles of mass $m$ at temperature $T$ in a force field
corresponding to some potential $V(\textbf{x})$. The most characteristic
implication of this assumption is that the space extension of the interaction
region increases with increasing temperature. This is the case for most models,
but it is not a law of nature. For instance, stars get hotter when they shrink.
The canonical distribution is being used in many models. The assumption of a
non-relativistic potential is not realistic, but it is sufficiently general to
reproduce any size and shape of the interaction region, so it seems
sufficiently flexible to provide qualitatively reliable results.

\section{The low- and high-temperature limits}

In the low temperature limit, classically, the particle rests
$(\textbf{p}=\textbf{0})$ at the minimum of the potential. Let us put there
$\textbf{x} = \textbf{0}$. Thus,

\begin{equation}\label{lowtcl}
  F(\textbf{p},\textbf{x}) = \delta(\textbf{x})\delta(\textbf{p}).
\end{equation}
Because of the sharp peak this cannot be interpreted as a Wigner function. In
order to get a candidate Wigner function, $F(\textbf{p},\textbf{x})$ must be
smeared.

The quantum-mechanical result is also easy to find. The particle must be in its
ground state. Denoting the corresponding wave function $\psi_0(\textbf{x})$ we
get

\begin{equation}\label{lowtqu}
W(\textbf{p},\textbf{x}) =
\int\!\!\frac{d^3y}{(2\pi)^3}\psi_0(\textbf{x}+\frac{\textbf{y}}{2})\psi_0(\textbf{x}
- \frac{\textbf{y}}{2})e^{-i\textbf{py}}.
\end{equation}
Now both position and momentum are spread around the point
$\textbf{p}=\textbf{0},\textbf{x}=\textbf{0}$. In the theory of fluctuations
this effect is referred to as quantum fluctuations. Formula (\ref{lowtqu}) can
be obtained by smearing (\ref{lowtcl}). Therefore, smearing can be interpreted
as a way of introducing quantum fluctuations. However, for each potential a
different smearing prescription would be needed. Thus, at low temperatures the
predictive power of the recipe: start with the classical distribution and smear
it, is poor.

According to our assumptions, the classical distribution is in general

\begin{equation}\label{maxbol}
  F(\textbf{p},\textbf{x} ) = Ne^{-\frac{\beta \textbf{p}^2}{2m} -\beta V(\textbf{x})}.
\end{equation}
Here and in the following $N$ denotes a normalization factor. Where the
normalization factors are of no interest, we will use the same notation for all
of them. The quantum-mechanical density operator is

\begin{equation}\label{rhoqua}
  \hat{\rho} = Ne^{-\frac{\beta \hat{\textbf{p}^2}}{2m} -\beta
  V(\hat{\textbf{x}})}.
\end{equation}
The difference between the classical and the quantum-mechanical expressions is
that in the latter the kinetic energy does not commute with the potential
energy. Let us note, however, that in the high-temperature limit $\beta$ tends
to zero. The commutator of the potential energy  and kinetic energy terms in
the exponent of (\ref{rhoqua}) is proportional to $\beta^2$ and, therefore, is
negligible. Accordingly, in the high-temperature limit the two description are
equivalent, as will be demonstrated more rigorously latter.

The results from this section correspond to an effect which is well known from
statistical physics. In the high-temperature limit the thermal fluctuations,
common to the classical and quantum descriptions, usually dominate while in the
low-temperature limit the quantum mechanical fluctuations, absent in the
classical case, are the important ones.

\section{The smearing density operators}

When a classical phase-space distribution violates the bound (\ref{wigbou}), it
is necessary to smear it. A way of doing this, is to introduce a (smearing)
density operator as close as possible to the classical density distribution.
Once a density operator is given, one can calculate from it a Wigner function
which satisfies all the consistency conditions.

Let us try, as smearing density operator, the operator

\begin{equation}\label{kinrho}
  \hat{\rho}_{sm} = N e^{\frac{\beta \hat{\textbf{p}}^2}{4m}} e^{-\beta V(\hat{\textbf{x}})}e^{\frac{\beta
  \hat{\textbf{p}}^2}{4m}}.
\end{equation}
The kinetic energy term has been split in order to make this operator
hermitian, as it should. The corresponding density matrix in the momentum
representation is

\begin{equation}\label{kinsme}
  \langle \textbf{p}|\hat{\rho}_{int}|\textbf{p}' \rangle = Ne^{-\frac{\beta}{2m}(\textbf{K}^2 +
  \frac{1}{4}\textbf{q}^2)}\int\!\!dx\;e^{-\beta V(\textbf{x}) - i\textbf{qx}},
\end{equation}
and for the Wigner function one gets

\begin{equation}\label{}
  W_{sm}(\textbf{K},\textbf{X}) = N e^{-\frac{\beta \textbf{K}^2}{2m}}\int\!\!dx\;e^{-\beta
  V(\textbf{x})}
  e^{-\frac{2m}{\beta}(\textbf{X} - \textbf{x})^2}.
\end{equation}
In the high temperature limit, $\beta \rightarrow 0$, the second exponent in
the integrand, taken with a suitable part of the normalizing factor, tends to
$\delta(\textbf{X} - \textbf{x})$ and, after integrating over $\textbf{x}$, one
obtains the classical density distribution. For low temperatures, of course, no
Wigner function can reproduce the classical distribution.

Let us consider as example the harmonic oscillator with $V(\textbf{x}) =
\frac{1}{2}k \textbf{x}^2$. One gets

\begin{equation}\label{oscgen}
  W_{sm}(\textbf{p},\textbf{x}) = N e^{-\frac{\beta \textbf{p}^2}{2m^*} - \frac{\beta}{2}k^* \textbf{x}^2},
\end{equation}
where

\begin{equation}\label{}
  m^* = m, \qquad k^* = \frac{1}{1 + \eta^2}k,
\end{equation}
and
\begin{equation}\label{}
  \eta = \frac{1}{2}\beta\omega.
\end{equation}
The parameter $\omega = \sqrt{\frac{k}{m}}$ is the frequency of the oscillator.
Note that with this smearing prescription the effective frequency $\omega_{eff}
= \sqrt{\frac{k^*}{m^*}}$ depends on temperature. In order to get after
smearing a distribution identical with $F(\textbf{p},\textbf{x})$ one would
have to make before smearing the substitution
\begin{equation}\label{ksubst}
  k \rightarrow (1 + \eta^2)k.
\end{equation}

At high temperatures $k^* \approx k$ and the classical result is reproduced. At
low temperatures, when $\eta$ is large, $k^* \ll k$ and the $x$-distribution is
smeared which avoids the contradiction to Heisenberg's uncertainty principle.

As another example let us choose

\begin{equation}\label{}
  \hat{\rho}_{sm} = N e^{-\frac{1}{2}\beta V(\hat{\textbf{x}})}e^{-\frac{\beta \hat{\textbf{p}}^2}{2m}}e^{-\frac{1}{2}\beta
  V(\hat{\textbf{x}})}.
\end{equation}
For the harmonic oscillator this yields again formula (\ref{oscgen}), but with

\begin{equation}\label{}
m^* = \left(1 + \eta^2\right) m,\qquad k^* = k.
\end{equation}
This time the smearing is in momentum space. A popular smearing prescription
\cite{KOP}, \cite{GKW} is

\begin{equation}\label{}
  \hat{\rho}_{sm} =
  \int\!\!dpdx\;F(\textbf{p},\textbf{x},t)|\psi(\textbf{p},\textbf{x})\rangle\langle\psi(\textbf{p},\textbf{x})|,
\end{equation}
where the state vector $|\psi(p,x)\rangle$ represents a bound state of one
particle with positions close to $\textbf{x}$ and momenta close to
$\textbf{p}$.

It is seen that various choices of $\hat{\rho}_{sm}$ correspond to various
smearing prescriptions. Each of them gives a reasonable Wigner function, but
only with (\ref{rhoqua}) chosen as the smearing density operator the correct
Wigner function is obtained. We will compare now in the general case the
smearing density operator (\ref{kinrho}) with the exact one (\ref{rhoqua}).

\section{Effective Hamiltonian}

The results obtained in the preceding section for the harmonic oscillator can
be generalized to other potentials. One always finds that the smeared Wigner
function corresponds to some Hamiltonian, but in general not to the true one
for the system being studied. We will call this Hamiltonian effective
Hamiltonian. It is defined by the relation

\begin{equation}\label{}
  \hat{\rho}_{sm} = Ne^{-\beta \hat{H}_{eff}}.
\end{equation}
The smearing density operator (\ref{kinrho}) has the form

\begin{equation}\label{}
  \hat{\rho}_{sm} = Ne^{\hat{X}}e^{\hat{Y}}e^{\hat{X}},
\end{equation}
Thus, ignoring the irrelevant constants $\log N$,

\begin{equation}\label{defhef}
  -\beta\hat{H}_{eff} = \log\left(e^{\hat{X}}e^{\hat{Y}}e^{\hat{X}}\right).
\end{equation}
The right-hand side can be evaluated from a simple extension of the famous
Baker-Campbell-Hausdorff formula. The result is a series, in general infinite,
of iterated commutators constructed from the operators $\hat{X}$ and $\hat{Y}$.
Since both $\hat{X}$ and $\hat{Y}$ are proportional to $\beta$, this is a power
series expansion in $\beta$. An elegant and convenient method for calculating
the coefficients of this series for a more general case, i.e. for
$\log\left(e^{\hat{X}}e^{\hat{Y}}e^{\hat{Z}}\right)$, has been described in
ref. \cite{REI}. In our case an additional simplification occurs. Note that the
operator $e^{-\hat{X}}e^{-\hat{Y}}e^{-\hat{X}}$ is the inverse of the operator
$e^{\hat{X}}e^{\hat{Y}}e^{\hat{X}}$. Therefore, its logarithm equals
$+\beta\hat{H}_{eff}$. On the other hand, the expansion of this logarithm can
be obtained by taking the expansion for (\ref{defhef}) and changing the signs
of all the $\hat{X}$-s and $\hat{Y}$-s. These two prescription are consistent
if and only if all the commutators with even numbers of factors have
coefficients zero. For instance, for the smearing density operator
(\ref{kinrho}) one finds

\begin{equation}\label{}
  \hat{H}_{eff} = \hat{H} - \frac{\beta^2}{6}\left[\hat{H},\left[\frac{\hat{\textbf{p}}^2}{4m},V(\hat{\textbf{x}})\right]
  \right] + \cdots,
\end{equation}
where $\hat{H} = \frac{\hat{p}^2}{2m} + V(\hat{x})$ is the original
Hamiltonian. The contribution of the single commutator vanishes as it should.

In particular, for the harmonic oscillator

\begin{equation}\label{}
  \hat{H}_{eff} = \hat{H} + \frac{\eta^2}{3}\left(\frac{\hat{\textbf{p}}^2}{2m} - k \hat{\textbf{x}}^2\right) +
  \cdots.
\end{equation}
Using a program in \textit{MATHEMATICA} given in \cite{REI} it is easy to
calculate more terms of this series. In order to get an effective Hamiltonian
corresponding to the original phase-space density, and not to its smeared
version, one must make in the Hamiltonian on the right-hand side the
substitution (\ref{ksubst}). This yields the Hamiltonian

\begin{equation}\label{hastar}
  \hat{H}^* = \left(1 + \frac{1}{3}\eta^2\right)\hat{H},
\end{equation}
which reproduces, to second order in $\eta$ the classical distribution
(\ref{maxbol}).

For the harmonic oscillator, it is easy to compare directly, without using a
smearing density operator, the Wigner function with the corresponding classical
distribution. This is discussed in the following section.

\section{Classical density and Wigner function for the harmonic oscillator}

For a harmonic oscillator at temperature $T$, the Wigner function, or
equivalently the density matrix, has been calculated by a variety of methods
\cite{IMR}, \cite{HIL}, \cite{ZAL2}. The result is

\begin{equation}\label{wigdis}
  W(\textbf{p},\textbf{x}) = N e^{-\beta\frac{tanh\eta}{\eta}( \frac{\textbf{p}^2}{2m}+
  \frac{1}{2}k\textbf{x}^2 )}.
\end{equation}
This is to be compared with the corresponding classical density

\begin{equation}\label{kladis}
  F(\textbf{p},\textbf{x}) = \left(\frac{\eta_{class}}{\pi}\right)^3e^{-\beta_{class}(\frac{\textbf{p}^2}{2m}+
  \frac{1}{2}k\textbf{x}^2 )}.
\end{equation}
According to condition (\ref{wigbou}), if $\eta_{class} \equiv
\frac{1}{2}\beta_{class}\omega > 1$ the classicl density must be smeared before
being interpreted as a Wigner function. The distributions (\ref{wigdis}) and
(\ref{kladis}) coincide, if

\begin{equation}\label{}
  \eta_{class} = \mbox{tanh}\;\eta.
\end{equation}
At high temperatures, where $\eta$ and $\eta_{class}$ are both small,
$\eta_{class} \approx \eta$ and there is no harm in interpreting the classical
distribution as a Wigner function. At low temperatures, however, $\eta$ can be
arbitrarily large, while $\eta_{class}$ never exceeds unity. Then, interpreting
the classical distribution as a Wigner function can lead to serious errors.

An obvious question is, where on this scale are situated the temperatures in
the range of some $(100-200)$ MeV relevant for high energy scattering? The
difficulty is that, what matters is the temperature in units of $\omega$, and
$\omega$ is not known. In order to get a rough estimate, let us make the
admittedly crude assumption that the results for the harmonic oscillator can be
used as a guide. For the harmonic oscillator

\begin{equation}\label{}
  \sigma^2(p_i) =\frac{\sqrt{km}}{2\;\mbox{tanh}\;\eta}, \qquad \sigma^2(x_i) =
  \frac{1}{2\;\sqrt{km}\;\mbox{tanh}\;\eta},\qquad i=1,2,3.
\end{equation}
This yields

\begin{equation}\label{}
  \mbox{tanh}\eta = \frac{1}{2\sqrt{\sigma^2(x_i)}\sqrt{\sigma^2(p_i)}}.
\end{equation}
Choosing a value typical for high energy scattering, $\sqrt{\sigma^2(p_i)} =
300$MeV, one gets as an approximation, which is very good when $|\eta -
\eta_{class}|$ is small,

\begin{equation}\label{temcor}
  T =
  \left(1 - \frac{0.036}{\sigma^2(x_i)}\right)
  T_{class},
\end{equation}
where $\sigma^2(x_i)$ should be expressed in squared fermis. It is seen that
for heavy ion high energy scattering, where typically $\sqrt{\sigma^2(x_i)}
\approx 5$fm, the correction is negligible. For $e^+e^-$ annihilations,
however, where $\sqrt{\sigma^2(x_i)}$ can be as small as $0.7$fm, the
correction is about seven percent. We conclude that, interpreting the classical
distribution as a Wigner function one always finds that the classical
temperature is higher than the one corresponding to the Wigner function
interpretation. Qualitatively this conclusion seems unavoidable. The quantum
fluctuations are not included in the classical description. In order to
reproduce their effect it is necessary to increase the thermal fluctuations,
which means increasing the temperature. The corresponding correction is
probably negligible for heavy ion collisions, but may be at the few percent
level for $e^+e^-$ annihilations.

In order to obtain a Wigner function of the form (\ref{maxbol}) with
$\beta_{class} = \beta$, one has to start with the Hamiltonian

\begin{equation}\label{}
  \hat{H}^* = \frac{\eta}{\mbox{tanh}\eta}\hat{H} = \left(1 + \frac{1}{3}\eta^2 +
  \cdots\right)\hat{H},
\end{equation}
which agrees with (\ref{hastar}) to second order in $\eta$.

\section{Conclusions}

Numerous models provide classical phase space distributions for the particles
produced in multiparticle production processes. When describing Bose-Einstein
correlations these densities, sometimes smeared, are being used as if they were
Wigner functions. Therefore, it is an interesting question how close, in
situations encountered in particle physics, are the classical phase space
distributions to their corresponding Wigner functions.

Converting a classical phase space distribution to a Wigner function, when
temperature is not very high, one should in principle consider quantum
fluctuations. The simplest way is to assume that they are negligible. Our
discussion, based on the analogy with the harmonic oscillator, suggests that
this could be legitimate for high-energy heavy ion collisions, but probably not
for $e^+e^-$ annihilations.

In general, quantum fluctuations are negligible at high temperatures and
important at low temperatures. For a given potential this means that, they are
important when the interaction region is small, and unimportant when it is
large. For the specific model discussed in the present paper, the correction
goes like $R^{-2}$ as seen from formula (\ref{temcor}). The correction always
reduces the inferred temperature of the system.

For potentials more complicated than that of the harmonic oscillator, it is
convenient to perform the comparison of the classical phase-space distribution
with the corresponding Wigner function in two steps. First one introduces a
smearing density operator, which should provide a Wigner function easy to
compare with the classical distribution. This is equivalent to the familiar
smearing and yields a Wigner function which satisfies all the general
consistency conditions. It can be done in an infinity of ways. Three are
described in the text. The introduction of the smearing density operator is
equivalent to the introduction of an effective hamiltonian which yields the
same Wigner function as the smeared density operator. In the first two examples
discussed here, using the Baker-Campell-Hausdorff formula, one can obtain this
effective Hamiltonian as a power series in $\beta$. The leading term is the
true Hamiltonian which confirms that in the high temperature limit ($\beta
\rightarrow 0$) the crude estimate of the quantum fluctuations, as done by
introducing smearing, is good enough. This is implied by the fact that quantum
fluctuations are negligible. \vspace{0.5in}

                              \Large{\textbf{Acknowledgements}}\normalsize

The author thanks Mark Gorenstein and Krzysztof Redlich for helpful
discussions.

\end{document}